%% file: main.tex
\pdfoutput=1 
\documentclass[lettersize,journal]{IEEEtran}
\usepackage{amsmath,amsfonts}
\usepackage{algorithmic}
\usepackage{algorithm}
\usepackage{array}
\usepackage{textcomp}
\usepackage{stfloats}
\usepackage{url}
\usepackage{verbatim}
\usepackage{graphicx}
\usepackage{cite}
\usepackage{multirow}
\usepackage{makecell}
\usepackage{pifont}
\usepackage{caption}
\usepackage{subcaption}
\usepackage[most]{tcolorbox}
\usepackage{colortbl,hhline}
\usepackage[normalem]{ulem}
\usepackage{float}
\usepackage{fontawesome}
\usepackage{subfiles}
\usepackage{booktabs}
\usepackage{bm}
\usepackage{fontawesome}
\usepackage{threeparttable}
\usepackage{tabularx}

\newcolumntype{P}[1]{>{\centering\arraybackslash}p{#1}}

\usepackage[colorlinks,linkcolor=blue,anchorcolor=blue,citecolor=blue]{hyperref}

\begin{document}

\title{Practitioners' Expectations on Log Anomaly Detection}

\author{Xiaoxue Ma, Yishu Li, Jacky Keung, Xiao Yu, Huiqi Zou, Zhen Yang, Federica Sarro and Earl T. Barr
\thanks{Xiaoxue Ma and Yishu Li are with the Department of Electronic Engineering and Computer Science, Hong Kong Metropolitan University, Hong Kong, China. E-mail: kxma@hkmu.edu.hk, sliy@hkmu.edu.hk.}
\thanks{Jacky Keung is with the Department of Computer Science, City University of Hong Kong, Hong Kong, China. E-mail: jacky.keung@cityu.edu.hk.}
\thanks{Xiao Yu is with the State Key Laboratory of Blockchain and Data Security, Zhejiang University, Hangzhou, China. E-mail: xiaoyu\_cs@hotmail.com.}
\thanks{Huiqi Zou is with the Department of Computer Science, Johns Hopkins University, Baltimore, United States. E-mail: hzou11@jh.edu.}
\thanks{Zhen Yang is with the School of Computer Science and Technology, Shandong University, Shandong, China. E-mail: zhenyang@sdu.edu.cn.}
\thanks{Federica Sarro and Earl T. Barr are with the Department of Computer Science, University
College London, London, U.K. E-mail: f.sarro@ucl.ac.uk, e.barr@ucl.ac.uk.}
\thanks{Digital Object Identifier 10.1109/TSE.2024.XXXXXXXX}}


\markboth{IEEE Transactions on Software Engineering}%
{Ma \MakeLowercase{\textit{et al.}}: Practitioners' Expectations on Log Anomaly Detection}


\maketitle

\begin{abstract}
Log anomaly detection has become a common practice for software engineers to analyze software system behavior. Despite significant research efforts in log anomaly detection over the past decade, it remains unclear what are practitioners' expectations on log anomaly detection and whether current research meets their needs. To fill this gap, we conduct an empirical study, surveying 312 practitioners from 36 countries about their expectations on log anomaly detection. In particular, we investigate various factors influencing practitioners' willingness to adopt log anomaly detection tools. We then perform a literature review on log anomaly detection, focusing on publications in premier venues from 2014 to 2024, to compare practitioners' needs with the current state of research. Based on this comparison, we highlight the directions for researchers to focus on to develop log anomaly detection techniques that better meet practitioners' expectations.
\end{abstract}


\section{Introduction}
\subfile{Introduction.tex}

\section{Research Methodology}
\subfile{Methodology.tex}

\section{Results}
\subfile{Results.tex}

\section{Discussion}
\subfile{Discussion.tex}

\section{Related Work}
\subfile{Related_Work.tex}

\section{Conclusion}
\subfile{Conclusion.tex}


\bibliographystyle{IEEEtran}
\bibliography{ref}

\newpage

 




\vfill

\end{document}

%% file: Introduction.tex
Logs are a crucial source of information in software systems, capturing system runtime behavior and aiding in software engineering tasks like program comprehension \cite{german2006using,ding2023temporal}, anomaly detection \cite{le2021log,meng2019loganomaly,lee2023heterogeneous,du2017deeplog}, and failure diagnosis \cite{chen2021pathidea,zhou2019latent}.
Extensive machine learning and deep learning-based techniques have been proposed in research for automated log anomaly detection \cite{yu2024deep,le2022log}, with the aim of reducing manual analysis costs. Concurrently, industrial log monitoring tools have emerged to provide log analysis services, including log anomaly detection.  
However, no prior studies have investigated practitioners' expectations of these techniques or tools. 
It remains unclear whether practitioners appreciate the current log anomaly detection techniques or log monitoring tools, the factors influencing their decisions to adopt them, and their minimum thresholds for adoption. Gaining insights from practitioners is essential to identify critical issues and guide researchers in developing solutions tailored to meet the needs of practitioners.

In this paper, we follow a mixed-methods approach to gain insights into practitioners' expectations on log anomaly detection: 1) Initially, we conduct a series of semi-structured interviews with 15 professionals with an average of 8.07 years of software development/maintenance experience. Through these interviews, we investigate the current use of log monitoring tools, the challenges associated with them, the perceived importance of automated log anomaly detection tools, and practitioners' expectations for these automated tools.
2) We then proceed with an exploratory survey involving 312 software practitioners from 36 countries. This quantitative approach allows us to validate practitioners' expectations uncovered in our interviews on a broader scale. 
3) Finally, we perform a comprehensive literature review on log anomaly detection spanning from 2014 to 2024, encompassing the last decade. We scrutinize research papers published in premier venues during this period, comparing these proposed techniques against the criteria that practitioners have for adoption.

We address the following four Research Questions (RQs):

\textbf{RQ1: What is the state of log monitoring tools, and what are the issues? }
The primary reasons surveyed practitioners abstain from using these tools include concerns or doubts about these tools, leading them to rely on manual analysis instead, and a lack of awareness regarding the existence of these tools.
About half of those with experience using these tools express dissatisfaction, citing issues such as ``The tool requires compatibility with different platforms and technologies'', ``The tool cannot provide a rationale for why a log is labeled as an anomaly'', and ``The tool cannot efficiently analyze large volumes of log data while maintaining effectiveness and efficiency''. Additionally, one-third of practitioners indicate such tools are unable to automatically detect log anomalies.
Furthermore, over 74\% of practitioners indicate that data resources such as historical labeled log data, metrics (system performance indicators), and traces (records of request journeys through a system) are sometimes or always available for log anomaly detection.

\textbf{RQ2: Are automated log anomaly detection tools important for practitioners? }
95.5\% of surveyed practitioners consider such automated tools to be essential or worthwhile for their software maintenance. Furthermore, many believe that user-friendly automated tools could enhance the effectiveness and efficiency of log anomaly detection, thus reducing the need for manual effort.

\textbf{RQ3: What are practitioners' expectations of automated log anomaly detection tools?} 
In the research area, there are two primary granularities for log aonmlay detection: log event level (a single log) and log sequence level (a log sequence consisting of multiple logs). Among the surveyed practitioners, 70.5\% tend to perform log sequence level analysis, 
with their first preference being to group log sequences according to window sizes.
Practitioners consider recall (identifying real anomalies), precision (accuracy of identified anomalies), and the efficiency of real-time anomaly detection as the most crucial evaluation metrics influencing their acceptance of these tools. Over 70\% expect recall and precision above 60\%. In addition, more than 78\% would consider using automated log anomaly detection tools if they could customize them to process diverse log structures and provide explanations for detected anomalies. 
Moreover, more than half of practitioners expect these tools to handle at least 100,000 logs, with installation, configuration, and learning taking no more than an hour. They also prefer anomalies to be identified within 5 seconds of their appearance.

\textbf{RQ4: How close are the current state-of-the-art log anomaly detection studies to satisfying practitioners' needs before adoption? } 
We identify 36 papers 
on log anomaly detection from 2014 to 2024, and explore the gap between proposed techniques and practitioners' expectations across nine aspects. 

Our main findings include: (1) Only 4 incorporate metrics or traces to aid in anomaly detection, despite 83.7\% and 74.9\% of practitioners indicating the availability of these data types.
(2) Only 6 papers explicitly perform log event level anomaly detection, which is the second preference for practitioners.
(3) Over half of the surveyed studies do not mention the log anomaly detection time, despite a majority advocating for real-time anomaly detection.
(4) Few or no studies address handling logs with diverse structures, providing a rationale for detected anomalies, customization, or privacy protection, all of which are significant concerns for surveyed practitioners. 
Despite most studies showcasing accurate anomaly detection on public datasets, half of the practitioners still refrain from adopting the proposed techniques due to main concerns over their interpretability, user-friendliness, and ability to handle various log data (details in Section~\ref{S:rq4}). This highlights a significant need for improvements to better meet practitioners' needs.

In summary, our work makes the following contributions:
\begin{itemize}
    \item We interview 15 professionals and survey 312 practitioners from 36 countries to gain insights into their expectations. This includes their perspectives on the current log monitoring tools, the importance of automated log anomaly detection tools, the factors influencing their adoption of these tools, and their minimum adoption thresholds. 

    \item We conduct an extensive literature review, following inclusion criteria and three steps of examination to identify relevant papers published in premier venues over the past decade. We then compare the current state of research with practitioners' expectations.

    \item We highlight potential implications to align research efforts with the needs of practitioners, such as integrating interpretable and generalizable techniques into automated tools and enhancing their user-friendliness and customization.
\end{itemize}

%% file: Methodology.tex
\begin{figure}[tb]
    \centering
      \includegraphics[width=1\linewidth]{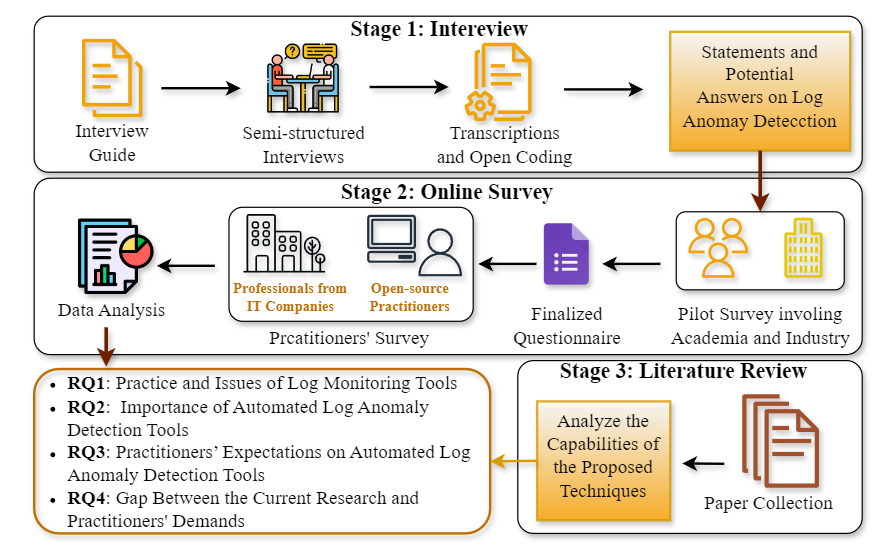}
      \captionsetup{skip=3pt}
      \caption{The overview of the research methodology.}
      \label{fig:Method}
\end{figure}

Our research methodology follows a mixed-methods approach \cite{leavy2022research}, as illustrated in Figure~\ref{fig:Method}, comprising three main stages. 
\textbf{\textit{Stage}} \textbf{1}: Interview with professionals to explore their practices in detecting log anomalies, their experiences, the issues they have encountered when using log monitoring tools for log anomaly detection, and their expectations on log anomaly detection.
\textbf{\textit{Stage}} \textbf{2}: Perform an online survey designed to validate and expand upon the findings derived from the interviews regarding log anomaly detection.
\textbf{\textit{Stage}} \textbf{3}: Conduct a comprehensive literature review to analyze whether or to what extent current state-of-the-art log anomaly detection techniques have fulfilled practitioners' needs and expectations. Both the interviews and survey comply with the policies of the relevant institutional review board.

\subsection{Stage 1: Interview}
\textbf{\textit{Protocol.}} 
We conduct a series of semi-structured interviews, utilizing both video calls and face-to-face discussions, to comprehensively examine practitioners' practices, issues, and expectations regarding log anomaly detection. Initially, we gather key points in log anomaly detection based on our experience with research articles and log monitoring tools to draft an interview guide, revising it as needed during the interview to ensure relevance.
Before each interview, we provide the interviewees with a brief introduction to the study's background information, ensuring they are informed about the recording of the interview, and emphasize that we will protect the practitioners’ identities. Each interview lasted for 40 to 60 minutes, and the first author presented to document the transcript and ask follow-up questions based on the interviewees' responses to obtain a more in-depth understanding.
In each interview, we first pose some demographic questions to the interviewees to ask about their background, including job roles and working experience. Subsequently, we explore their practices with log monitoring tools for log anomaly detection and the issues they encountered. Finally, we ask open-ended questions to gather their perceptions and expectations regarding log anomaly detection.

\textbf{\textit{Interviewees.}}
We invite 15 interviewees from eight IT companies worldwide, including Microsoft, Google, Alibaba, Intel, Huawei, etc. 
These interviewees are professionals from various roles closely engaged in log analysis within software development and maintenance, such as developers, operation and maintenance engineers, and others. The IT project experience of our interviewees varies from 4 years to 16 years, with an average professional experience of 8.07 years (minimum: 4, median: 7, maximum: 16, standard deviation: 3.45 years).

\textbf{\textit{Open Coding Analysis.}}
The first author transcribes the interviews and performs open coding \cite{spencer2009card} to generate an initial set of codes. The second author then verifies the codes and provides suggestions for improvement. After incorporating the suggestions, the two authors independently analyze and sort the opinion cards into potential descriptions for the questionnaire. 
The Cohen’s Kappa value between the two authors is 0.72, indicating a substantial agreement. Any disagreements have been discussed to reach a consensus. To mitigate the bias of the two authors in sorting descriptions, another two authors have also reviewed and confirmed the final set of survey descriptions. Ultimately, based on the results of the interviews, we identify eight issues with log monitoring tools for log anomaly detection, eight reasons for not using them, five degrees of importance regarding automated log anomaly detection tools, and nine expectations from these tools.

\subsection{Stage 2: Online Survey}
\subsubsection{Survey Design} 
The survey comprises various question types, including single-choice, multiple-choice, rating questions on a 5-point Likert scale, ranging from ``Strongly Disagree'' to ``Strongly Agree'', and short answer questions. Additionally, we include an ``I don't know'' option for surveyed practitioners who do not understand our descriptions. To mitigate bias resulting from practitioners' unfamiliarity with log anomaly detection, we provide a detailed explanation of its workflow and user scenarios.

\textbf{(1) Demographics.} This section collects demographic information of the surveyed practitioners, including their countries of residence, primary job roles, professional experiences and programming languages, and team sizes.  

\textbf{(2) Practices and Issues of Log Monitoring Tools for Log Anomaly Detection.} This section first offers surveyed practitioners a concise overview of log monitoring tools for detecting log anomalies. Then, it explores the specific tools they have employed and the issues they have met during usage for log anomaly detection. For practitioners who have not utilized such tools, we inquire about their primary reasons. In addition, we investigate the availability of data resources to aid in log anomaly detection. 

\textbf{(3) Importance of Automated Log Anomaly Detection Tools.} 
Given practitioners' potential doubts and concerns about existing log monitoring tools for log anomaly detection, or their possible unawareness of such tools, this section assesses the importance of an automated log anomaly detection tool, assuming it can meet practitioners' expectations. Practitioners are asked to evaluate the importance of the automated tool using statements such as ``Essential'' (I will use this tool daily), ``Worthwhile'' (I will use this tool), ``Unimportant'' (I will not use this tool), or ``Unwise'' (This tool will have a negative impact on log anomaly detection for me or my team). Accordingly, practitioners are queried about their primary motivations for using or reasons for not using the automated tool.

\textbf{(4) Practitioners’ Expectations on Automated Log Anomaly Detection Tools.} 
This section investigates surveyed practitioners' expectations regarding the detection granularity levels for utilizing automated log anomaly detection tools, including log event level (detecting a single log) and log sequence level (detecting a log sequence containing multiple logs). Strategies for grouping logs into log sequences include grouping logs based on window sizes (e.g., 2-20 logs per sequence), fixed time intervals (e.g., logs within every 5 minutes per sequence), and timestamps or sessions (e.g., block\_id). 
The section then explores the key factors influencing their acceptance of these tools for software system maintenance, such as \textit{correct detection rates} (whether the real anomalies can be correctly identified by the tools and whether the identified anomalies by the tools are real anomalies), \textit{real-time detection} (whether abnormal logs are detected in real-time as they occur), \textit{interpretability} (whether the tool can provide a rationale for why a detected log is labeled as an anomaly), \textit{generalizability} (whether the tool can analyze logs with various structures), \textit{customization} (whether the tool can be easily adjusted to meet different requirements, like adding new algorithms or adapting thresholds for defining anomalies), \textit{easy to use} (whether the tool is easy to install, configure, and utilize), and \textit{security and privacy measures} (whether privacy and security measures are implemented and stated in the tools for sensitive information protection).
Furthermore, the section explores the minimum adoption threshold of such tools in terms of effectiveness (measured by recall and precision values), efficiency (time required to detect anomalies and time from the tool installation to successful utilization), scalability (capacity to handle a specified number of logs), and privacy protection (required measures provided by the tool).

Considering many deep learning-based techniques in academia have been proposed for log anomaly detection and have demonstrated strong performance (i.e., recall and precision over 80\%) on public datasets, we ask practitioners about their willingness to use such techniques. Finally, practitioners are invited to provide free-text comments regarding automated log anomaly detection and our survey. In particular, we inquire about potential opportunities for leveraging large language models (LLMs) like ChatGPT to enhance log anomaly detection. Practitioners may or may not choose to provide final comments.
Before launching the survey, we conduct a pilot survey involving three industrial experts and two academics specializing in software operation and maintenance research, none of whom are interviewees or surveyed practitioners. This pilot aims to gather feedback on the survey's length, clarity, and the understandability of terms used. Based on their insights, we made minor adjustments to the draft survey, refining it into a finalized version. To cater to a global audience, we offer an English version of the survey through Google Form \cite{googleform}, with the full text publicly accessible \cite{survey}. 
Additionally, we translate the survey into Chinese and host it on a popular survey platform in China \cite{wjx} to ensure accessibility for practitioners in the region.

\subsubsection{Participant Recruitment} 
To gather a diverse pool of surveyed practitioners, we initially engage professionals within our social and professional circles employed at various IT companies. We request their support in sharing our survey with their colleagues. Invitations are extended to contacts at prominent companies such as Google, Microsoft, Intel, ByteDance, Cisco, Alibaba, Huawei, and others. Through this approach, we gather 76 responses. 
In addition, we contact contributors from GitHub repositories (via email addresses they made publicly available on GitHub), primarily those focused on log analysis. We prioritize repositories hosting popular open-source projects based on star ratings. Survey invitations are sent to 9,933 potential developers via email. Of these, 436 automatic replies indicate the recipients' absence, and 68 email addresses are unreachable. After excluding one incomplete survey and 17 responses with completion times of less than three minutes, as well as 2 responses from individuals with teaching roles and one with an unspecified role, we obtain 236 valid responses. In total, we receive 312 valid responses.
These responses originate from 36 countries, with China, the United States, and Germany among the top three regions with the highest number of surveyed practitioners. Most surveyed practitioners actively engage in software development (40\%) and operation and maintenance (29\%), with 3-10 years of experience in medium-sized project teams comprising 6 to 40 members. 


\subsubsection{Result Analysis} 
We analyze the survey results based on question types. For single-choice and multiple-choice questions, we report the percentage of each selected option. Open-ended questions undergo qualitative analysis through careful examination of responses. 
We draw bar charts to highlight possible trends in the Likert-scale answers. 
For the open-ended questions, the first two authors independently analyzed these, categorizing them into specific characteristics.
We exclude ``I don't know'' ratings since they constitute a small minority, i.e. less than 1\% of all valid responses.

\subsection{Stage 3: Literature Review}
First, we use IEEE, ACM, DBLP, and Google Scholar to search for publications from 2014 to 2024 by using the query: (``log''$||$ ``log anomaly'') \& (``detect/detection'' $||$ ``analyze/analysis'' $||$ ``predict/prediction'' $||$ ``classify/classification''). Then, we keep those papers published in premier peer-reviewed venues\footnote{We consider publications in conferences ranked A and A* according to the CORE Ranking (https://portal.core.edu.au/conf-ranks/) and journals falling in the Quartile Q1 according to the Journal Citation Reports (JCR) (https://jcr.clarivate.com/jcr/home).}, yielding 88 papers. 
From 88 papers, we then filter out irrelevant publications according to the following \textbf{inclusion criteria}: 1) The publication should propose a new log anomaly detection technique, and 2) the publication predicts a dichotomous outcome (i.e., normal or abnormal log). 
To assess whether the publications satisfy our inclusion criteria, we manually examine each publication following the three steps adopted in previous surveys \cite{hort2021survey, moussa2022use}: 1)\textbf{ Title:} if the publication's title clearly does not match our inclusion criteria, then it is excluded; 2) \textbf{Abstract:} if the publication's abstract does not meet our inclusion criteria, then it is excluded; and 3) \textbf{Body:} if the examined publication neither satisfies the inclusion criteria nor contributes to this survey, then it is excluded.
Among the 88 articles, 24 meet the inclusion criteria. 
Starting from these 24 articles, we perform one level of forward snowballing \cite{wohlin2014guidelines} and gather 71 additional publications. 
We assess these additional publications by using the same inclusion criteria and process described above, identifying 12 articles that meet the criteria.
Ultimately, we identify a total of 36 papers: 6 from ISSRE, 5 from ICSE, 3 from ESEC/FSE, 3 from ASE, 2 from KDD, 2 from ICWS, 1 from AAAI, 1 from CCS, 1 from ICDM, 1 from ACSAC, 1 from ICSME, 3 from ASEJ, 2 from TSC, 2 from TNSM, 1 from TOSEM, 1 from IJCAI, and 1 from ICT Express.

For each log anomaly detection paper, two authors read its content and analyze the capabilities of the proposed technique. For instance, Guo et al. \cite{guo2024logformer} utilize a Transformer-based framework, combining pre-training on the source domain and adapter-based tuning on the target domain, and evaluate it on three datasets with different log grouping strategies. Then we infer that they focus on log sequence level anomaly detection, supporting cross-project generalizability. 
We assess effectiveness and efficiency satisfaction rates using the lowest precision/recall values and the maximum detection time provided.
The first two authors discuss any differences in their capability analysis and confirm the final results through further reading of the papers.

%% file: Results.tex
\subsection{RQ1: Practices and Issues of Log Monitoring Tools}
In this research question, we investigate practitioners' practices regarding log monitoring tools, focusing on their tool preferences for log anomaly detection, reasons for non-utilization, data source availability, and encountered issues during tool usage for log anomaly detection.

\subsubsection{Practices of log monitoring tools} 
\begin{figure}[htbp]
  \centering
  \includegraphics[width=0.9\linewidth]{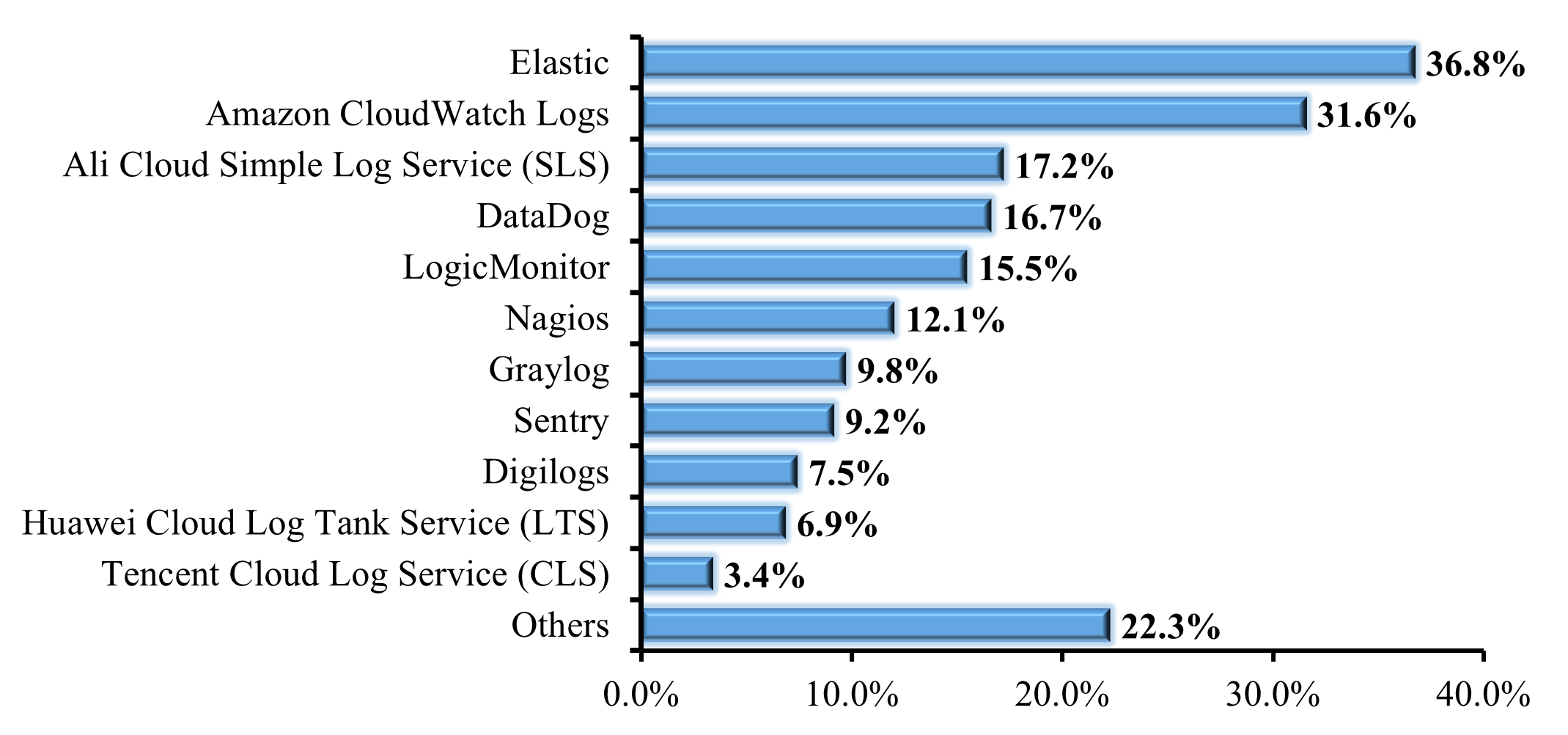}
  \captionsetup{skip=3pt}
  \caption{The percentages of tool usage for log anomaly detection.}
  \label{fig:ToolUsage}
\end{figure}

According to survey results, 174 out of 312 (55.8\%) practitioners report using log monitoring tools for log anomaly detection, while the remaining 138 indicate they do not use these tools. Figure~\ref{fig:ToolUsage} illustrates the percentage of usage of various tools for log anomaly detection. Practitioners may use more than one tool, resulting in a total percentage exceeding 100\%. The results reveal that 36.8\% of practitioners use \textit{Elastic}, and 31.6\% utilize \textit{Amazon CloudWatch Logs}. Following tools, \textit{Ali Cloud Simple Log Service}, \textit{DataDog}, \textit{LogicMonitor}, and \textit{Nagios} rank as the third to sixth most commonly used tools, with adoption rates of 17.2\%, 16.7\%, 15.5\%, and 12.1\%, respectively. Less prevalent tools, including \textit{Graylog}, \textit{Sentry}, \textit{Digilogs}, \textit{Huawei Cloud Log Tank Service}, and \textit{Tencent Cloud Log Service}, have adoption rates below 10\%. Tools with usage rates under 3\%, such as \textit{Better Stack}, \textit{LogCat}, and \textit{LogDNA}, along with tools specified by practitioners (e.g., internal systems), are categorized as \textit{Others}, with a total usage rate of 22.3\%.

Among practitioners who have used log monitoring tools, one-third of practitioners report the tools cannot automatically identify log anomalies and require manual analysis. We further investigate their awareness of the underlying techniques of these tools. The results indicate that 75.5\% of tools utilize keyword-based heuristic methods, while 26.3\% of tools are developed with machine learning or deep learning techniques. Some tools incorporate intersecting techniques, leading to a total percentage exceeding 100\%. 



\begin{center}
    \resizebox{\linewidth}{!}{
\begin{tabular}{l!{\vrule width 1pt}p{0.9\columnwidth}}
    \makecell{{\LARGE \faLightbulbO}}  &\textbf{Finding 1.} 
    Over half (55.8\%) of practitioners use log monitoring tools for log anomaly detection, with \textit{Elastic} and \textit{Amazon CloudWatch Logs} being the top choices. Notably, one-third find these tools incapable of automated anomaly detection.
\end{tabular}}
\end{center} 

For the 138 practitioners who have not used log monitoring tools for log anomaly detection, the primary reasons cited are that manual analysis suffices for detecting abnormal logs (49.3\%) and a lack of awareness of existing tools (47.1\%). Additionally, other reasons mentioned due to the multiple-choice nature of the question include a lack of technical expertise to use the tools (29\%), doubts about the effectiveness and reliability of the tools (21\%), a lack of compatibility with different platforms and technologies (15.9\%), time-consuming installation and learning process of the tools (13.8\%), data security or user privacy concerns regarding using the tools (13\%), expensive payment for using the tools (4\%). As some of the surveyed practitioners noted:

\noindent \faPencilSquareO \ \textit{Manual log analysis is straightforward and familiar to me. I haven't found an automated tool that can replicate our daily practices as effectively as I'd like.}

\noindent \faPencilSquareO \ \textit{Our business is intricate, and we’re skeptical about whether the automated tools can really adapt to our needs for finding actual anomalies. That’s why we’re still relying on manual analysis.}


\noindent \faPencilSquareO \ \textit{I wasn't even aware that such tools existed. If I had known about them, I would definitely consider giving them a try, especially because the manual workload can be quite heavy.}

To summarize, apart from the unawareness of existing log monitoring tools, most practitioners still have concerns or doubts about the performance of these tools in multiple aspects.


\begin{center}
    \resizebox{\linewidth}{!}{
\begin{tabular}{l!{\vrule width 1pt}p{0.9\columnwidth}}
    \makecell{{\LARGE \faLightbulbO}}  &\textbf{Finding 2.} 
Nearly half of the surveyed practitioners refrain from using log monitoring tools, primarily due to either their reliance on manual analysis, driven by concerns and doubts about these tools, or their unawareness of the existence of such tools.
\end{tabular}}
\end{center}

\begin{figure}[htbp]
  \centering
  \includegraphics[width=1\linewidth]{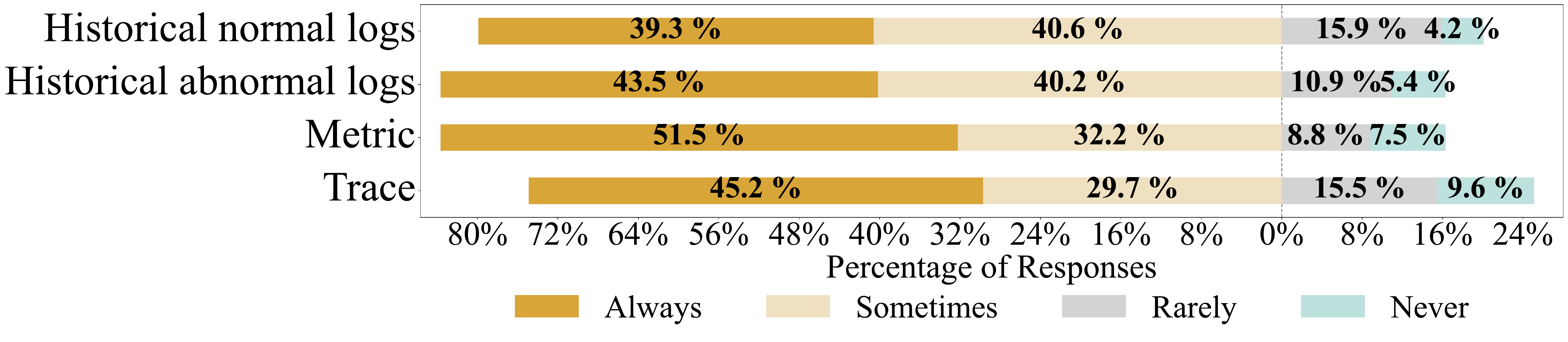}
  \captionsetup{skip=3pt}
  \caption{Data resource availability for log anomaly detection.}
  \label{fig:DataAvail}
\end{figure}

Figure~\ref{fig:DataAvail} illustrates the availability of data resources utilized by the surveyed practitioners in detecting log anomalies. They rate the availability of four specific data types (i.e., historical labeled normal logs, historical labeled abnormal logs, metrics, and traces) on a scale from ``Always'' to ``Never''. 
Metrics represent numerical data points indicating system performance indicators like response time, CPU usage, and memory consumption, while traces provide a comprehensive record of request journeys through a system, detailing timestamps, resource usage, and module interactions.
The survey indicates that these resources are \textit{always} available to 39.3\%, 43.5\%, 51.5\%, and 45.2\% of practitioners, respectively. Furthermore, 40.6\%, 40.2\%, 32.2\%, and 29.7\% report these data sources as \textit{sometimes} available. Less than 20\% of practitioners indicate that historical labeled normal and abnormal log data, as well as metrics, are \textit{rarely} or \textit{never} available for log anomaly detection. 
This suggests that while historical log data is available to more than 80\% of practitioners, over 74\% recognize the availability of other data types, such as metrics and traces, for log anomaly detection, as highlighted by one practitioner:

\noindent \faPencilSquareO \ \textit{I usually use the metric and trace data because they contain more useful information for detecting log anomalies.}

\begin{center}
    \resizebox{\linewidth}{!}{
\begin{tabular}{l!{\vrule width 1pt}p{0.9\columnwidth}}
    \makecell{{\LARGE \faLightbulbO}}  &\textbf{Finding 3.} 
  At least 74\% of the surveyed practitioners recognize the availability of historical labeled log data, metrics, and traces, with some indicating that these available data are helpful for log anomaly detection.
\end{tabular}}
\end{center}

\begin{figure*}[htbp]
  \centering
  \includegraphics[width=\linewidth]{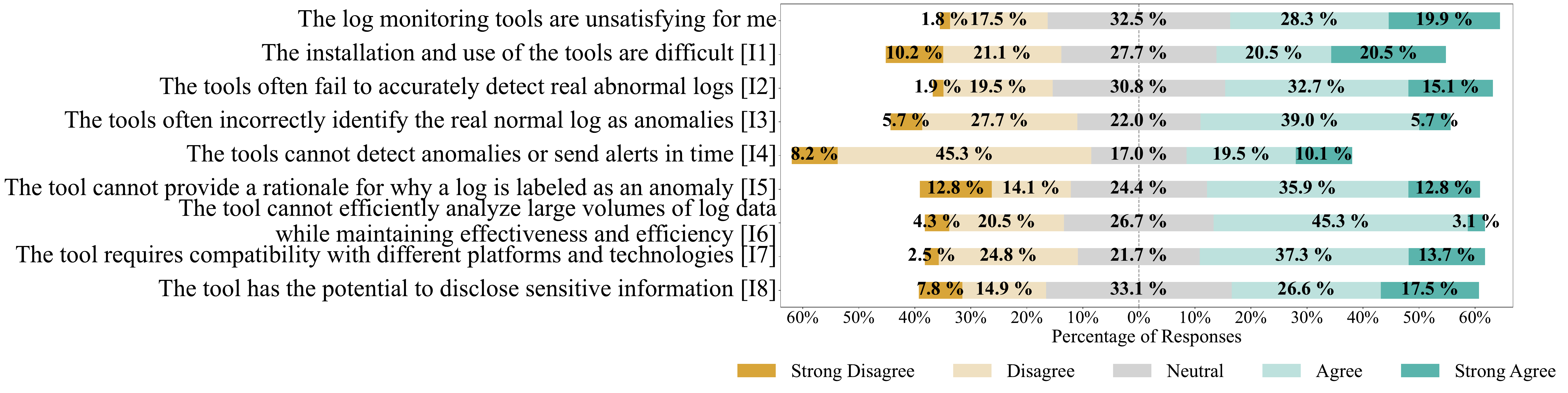}
  \captionsetup{skip=3pt}
  \caption{The current issues with log monitoring tools for log anomaly detection.}
  \label{fig:ToolIssues}
\end{figure*}

\subsubsection{Issues of log monitoring tools}~\label{subsec:issues}
Figure~\ref{fig:ToolIssues} demonstrates the issues encountered by surveyed practitioners when using log monitoring tools for log anomaly detection. Among practitioners utilizing these tools, 48.2\% express dissatisfaction, and only 19.3\% think these tools are satisfying. Over half of the practitioners (51\%) consider compatibility to be the most significant issue. They present agreement that ``the tool requires compatibility with different platforms and technologies''. 48.7\% of them consider ``the tool cannot provide a rationale for why a log is labeled as an anomaly''. Some practitioners share their opinions:

\noindent \faPencilSquareO \ \textit{The tool didn't fit well with our system, making it challenging to incorporate it into our workflow.}

\noindent \faPencilSquareO \ \textit{The tool is very simple. It just provides daily reports of anomalies that may be suspect and require much investigation to rule as a legitimate threat or nuisance scan.}

\noindent \faPencilSquareO \ \textit{The anomalies detected lack sufficient details, making them less meaningful for our needs. That means we still rely on manual analysis to gather comprehensive information.}

Practitioners also express significant concern about the scalability of these tools. Specifically, 48.4\% of the practitioners report that ``the tool cannot efficiently analyze large volumes of log data while maintaining effectiveness and efficiency''. These concerns are primarily voiced by practitioners working on medium and large-sized product development teams.

\noindent \faPencilSquareO \ \textit{Sometimes, the tool is useful when handling logs during the development and testing phases. However, its accuracy and performance drop when dealing with logs from the production phase, where data usage is increasing.}

Another two main challenges pertain to the effectiveness of the log monitoring tools. A total of 47.8\% of practitioners indicate that ``the tools often fail to accurately detect real abnormal logs'', whereas 44.7\% complain ``the tools often incorrectly identify the real normal log as anomalies''. Some practitioners express doubts about the accuracy of these tools, leading to reduced reliance on them.

\noindent \faPencilSquareO \ \textit{It's frustrating when the tool incorrectly detects anomalies because then we have to spend extra time doing manual analysis. Because of that, I find myself relying on it less for anomaly detection.}

Other issues highlighted by practitioners encompass concerns regarding privacy, usability, and efficiency.
Notably, 44.1\% express concerns regarding ``the tool has the potential to disclose sensitive information'', while 41\% find ``the installation and use of the tools are difficult''. Additionally, 29.6\% report that ``the tools cannot detect anomalies or send alerts in time''.


\begin{center}
    \resizebox{\linewidth}{!}{
\begin{tabular}{l!{\vrule width 1pt}p{0.9\columnwidth}}
    \makecell{{\LARGE \faLightbulbO}}  &\textbf{Finding 4.} 
   Practitioners highlight several notable issues with existing log monitoring tools for log anomaly detection. The majority (51\%) express concerns about the compatibility of these tools with different platforms and technologies. Between 40\% and 50\% of practitioners raise concerns regarding various aspects of the tools, including their interpretability, scalability, effectiveness, privacy disclosure, and ease of use.
\end{tabular}}
\end{center}

\subsection{RQ2: The importance of automated log anomaly detection tools} 
In this research question, we investigate how practitioners evaluate the importance of automated log anomaly detection tools, assuming their expectations are met.
Figure~\ref{fig:ToolImportance} presents the ratings from surveyed practitioners. About 95.5\% of surveyed practitioners regard automated log anomaly detection tools as either \textit{essential} or \textit{worthwhile}. 
Comments left by practitioners indicate that their primary motivation for using these tools is the reduction of manual analysis effort. As some practitioners suggested:

\begin{figure}[htp]
\centering
  \includegraphics[width=1\linewidth]{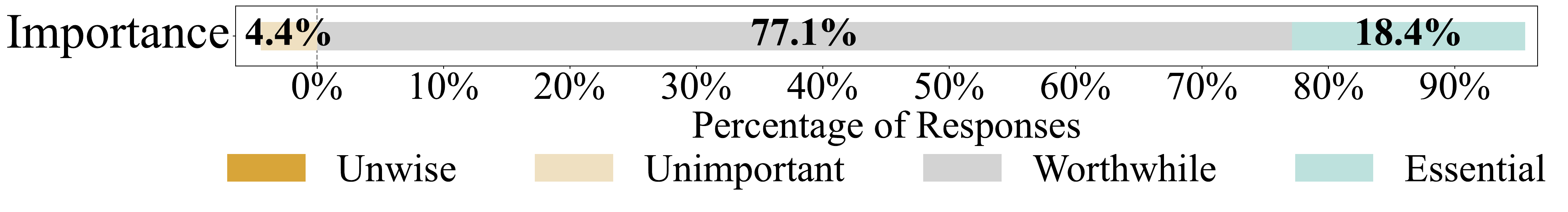}
  \captionsetup{skip=3pt}
  \caption{The importance of automated log anomaly detection tools.}
  \label{fig:ToolImportance}
\end{figure}

\noindent \faPencilSquareO \ \textit{If the tool can accurately detect log anomalies and provide detailed information, it would really cut down the time we spend on investigating and analyzing issues.}

\noindent \faPencilSquareO \ \textit{Having an automated tool is great; it really helps improve our work efficiency. I will definitely use it if the tool can provide accurate anomaly detection.}



A mere 4.4\% of practitioners regard the automated tool as \textit{unimportant} and indicate they would not use such tools. Their primary concerns are doubts about its effectiveness and a preference for internally developed tools over external ones.

\noindent \faPencilSquareO \ \textit{I doubt that the tool can be developed to adapt to our complex system and be easy to use.}

\noindent \faPencilSquareO \ \textit{Our product is designed for internal usage and will not be integrated with external log systems.}

\begin{center}
    \resizebox{\linewidth}{!}{
\begin{tabular}{l!{\vrule width 1pt}p{0.9\columnwidth}}
    \makecell{{\LARGE \faLightbulbO}}  &\textbf{Finding 5.} 
    The majority of practitioners (95.5\%) view automated log anomaly detection tools as either essential or worthwhile for their practice. They believe that such automated tools can efficiently and accurately detect anomalies, thereby streamlining manual analysis efforts.
\end{tabular}}
\end{center}

\subsection{RQ3: Practitioners’ Expectations on Automated Log Anomaly Detection Tools}\label{S:rq3}
In this research question, we delve into practitioners' expectations regarding log anomaly detection, exploring aspects including \textit{granularity level}, \textit{evaluation metrics}, \textit{effectiveness}, \textit{efficiency}, \textit{scalability}, and \textit{privacy protection}.

\begin{figure}[htbp]
  \centering
  \includegraphics[width=1\linewidth]{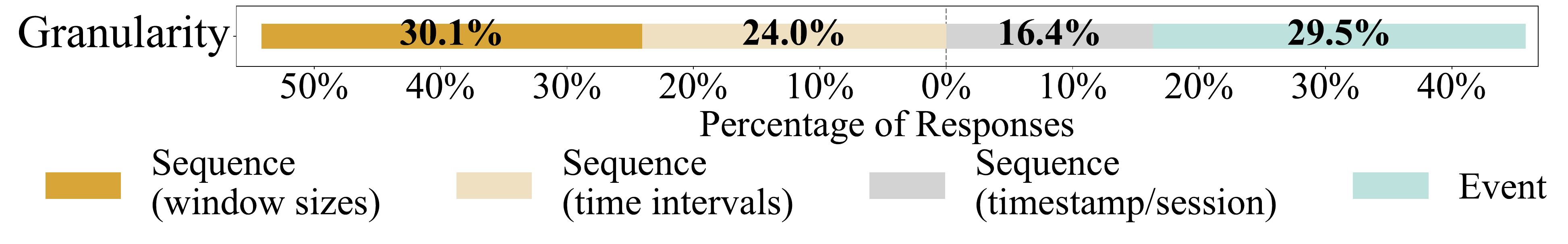}
  \captionsetup{skip=3pt}
  \caption{The granularity of automated log anomaly detection tools.}
  \label{fig:Granularity}
\end{figure}

\textbf{Granularity level.}
From Figure~\ref{fig:Granularity}, 70.5\% of practitioners prefer detecting log anomalies at the sequence level, where an entire sequence is labeled as abnormal if any log within it is abnormal. Specifically, 30.1\% group logs into sequences based on window sizes (e.g., every 20 logs), 24.0\% use fixed time intervals (e.g., every 5 minutes), and 16.4\% group logs by timestamp/session (e.g., block\_id). Furthermore, 29.5\% prefer log event level, making it the second most favored detection granularity among surveyed practitioners.

\begin{figure}[!htbp]
  \centering
  \includegraphics[width=1\linewidth]{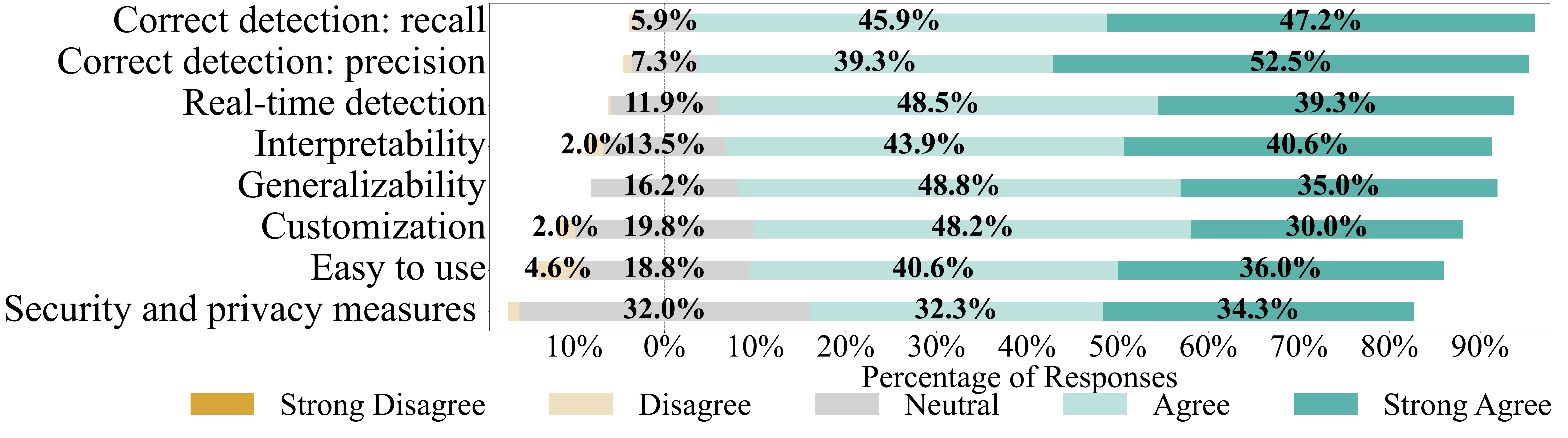}
  \captionsetup{skip=3pt}
  \caption{The factors affecting practitioners' acceptance of
using automated log anomaly detection tools.}
  \label{fig:FactorsForUsing}
\end{figure}

\textbf{Evaluation metrics.}
We analyze surveyed practitioners' opinions on evaluating automated log anomaly detection tools as shown in Figure~\ref{fig:FactorsForUsing}. The findings reveal that 93.1\% and 91.8\% of practitioners agree that the ability to correctly identify real anomalies and the accuracy of the identified anomalies (\textbf{\textit{correct detection rates}}) are the two most preferred evaluation metrics. 
As one practitioner noted: ``\textit{I want the tool to be able to detect potential anomalies and miss as few as possible}'' while another stated: ``\textit{I only accept the tool if it can detect anomalies with high accuracy; otherwise, I'll have to manually judge, costing me time.}''.
Additionally, 87.8\% of practitioners emphasize the importance of \textbf{\textit{real-time detection}} for identifying anomalies. As a practitioner noted: ``\textit{I'm really looking for a real-time tool that can streamline the log anomaly detection process, making it much easier and ultimately boosting the detection efficiency.}''.
Furthermore, 84.5\% value \textbf{\textit{interpretability}}, indicating that the tool should provide a rationale for why a log is labeled as an anomaly. 
\textbf{\textit{Generalizability}} is crucial for 83.8\% of practitioners, who think the tool should effectively analyze logs with diverse structures. As a practitioner emphasized: ``\textit{Detecting anomalies shouldn't always need specific knowledge of certain applications. It'd be great to automatically spot different log anomalies for wider use.}''.
\textbf{\textit{Customization}} is also important, with 78.2\% believing the tool should be easily adjustable to meet various requirements. For example, they want the ability to select different anomaly detection algorithms, configure specific log formats, and set custom alert thresholds.
Moreover, 76.6\% of practitioners consider \textbf{\textit{easy to use (including installation and deployment)}} as a significant factor for adopting the tool. 
One practitioner also emphasizes that if the tool effectively fulfills its purpose, the time required for installation and configuration is acceptable: ``\textit{If a tool gets the job done, the installation and configuration process is just part of the deal. Honestly, overcoming a challenge during setup can even make the tool more interesting to use. What really matters is that it does what it's supposed to do.}''.
Lastly, 66.6\% of practitioners agree that \textbf{\textit{security and privacy measures}} should be implemented to protect sensitive information.

\begin{center}
    \resizebox{\linewidth}{!}{
\begin{tabular}{l!{\vrule width 1pt}p{0.9\columnwidth}}
    \makecell{{\LARGE \faLightbulbO}}  &\textbf{Finding 6.} 
    More than 78\% of surveyed practitioners consider using automated log anomaly detection tools if they can be customized to process logs with different structures and provide a rationale for the detected anomalies.
\end{tabular}}
\end{center} 

\begin{figure}[!htbp]

  \centering
  \includegraphics[width=1\linewidth]{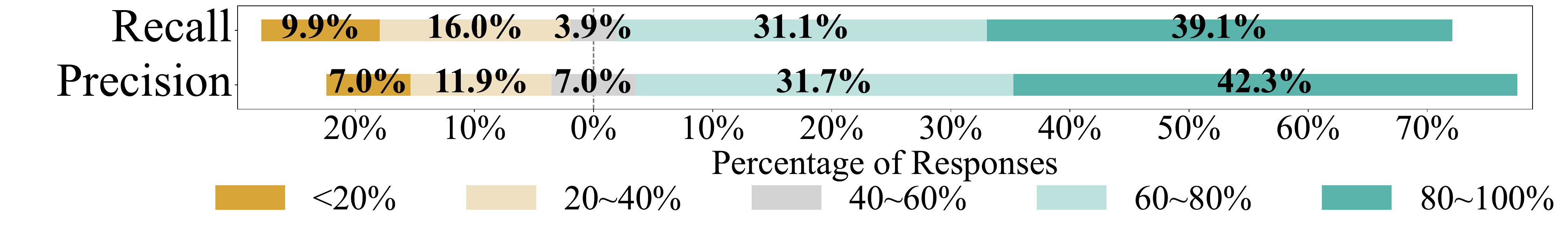}
  \captionsetup{skip=3pt}
  \caption{The surveyed practitioners’ satisfaction rate in terms of effectiveness.}
  \label{fig:ExpectEffectiveness}
\end{figure}

\textbf{Effectiveness.}
Figure~\ref{fig:ExpectEffectiveness} illustrates practitioners' satisfaction levels with automated log anomaly detection tools based on recall and precision values. The satisfaction rate is categorized into five capability ranges, from less than 20\% to 80\%-100\%. 
Our survey indicates that 39.1\% of practitioners will accept the tool only if the recall exceeds 80\%, and 42.3\% will do so if the precision exceeds 80\%. Achieving recall and precision above 60\% will satisfy 70.2\% and 74.0\% of practitioners, respectively.


\begin{center}
    \resizebox{\linewidth}{!}{
\begin{tabular}{l!{\vrule width 1pt}p{0.9\columnwidth}}
    \makecell{{\LARGE \faLightbulbO}}  &\textbf{Finding 7.} 
    Correct detection rates (recall and precision) are the most critical factors influencing surveyed practitioners' acceptance of automated log anomaly detection tools. Over 70\% of the surveyed practitioners expect the correct detection rates to exceed 60\%.
\end{tabular}}
\end{center}

\begin{figure}[htbp]
    \centering
    \includegraphics[width=\linewidth]{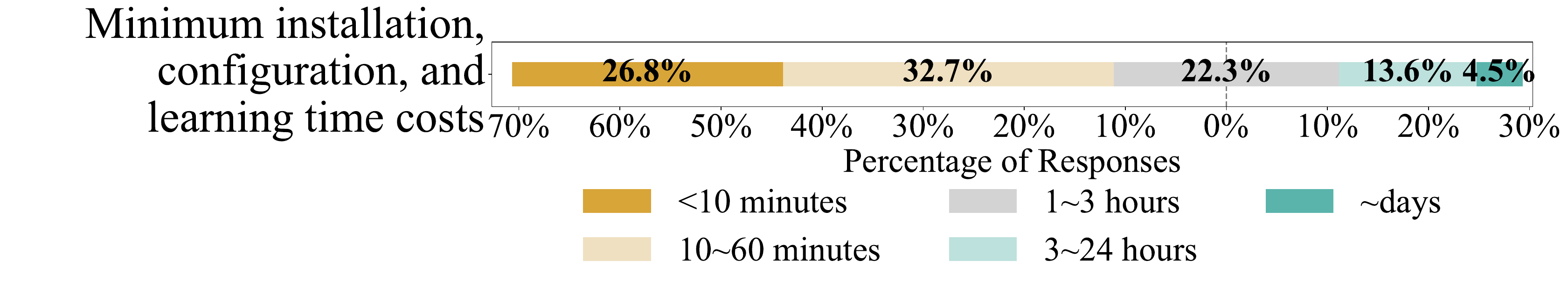}
    \includegraphics[width=\linewidth]{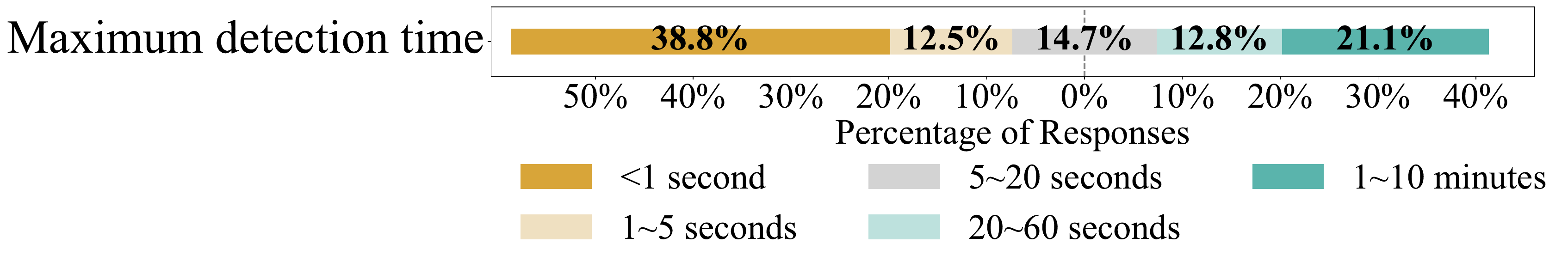}
    \includegraphics[width=\linewidth]{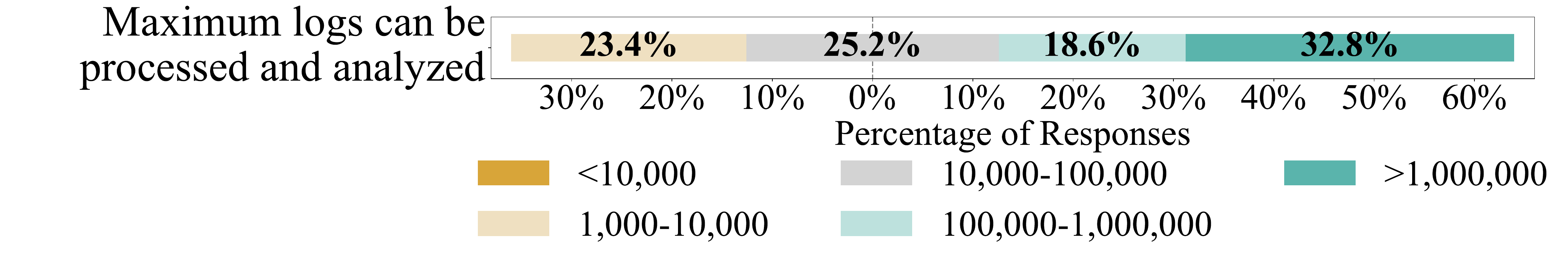}
    \captionsetup{skip=3pt}
    \caption{The surveyed practitioners' satisfaction rate with various capability ranges in terms of efficiency and scalability.}
    \label{fig:ExpectScal+Eff+LearnTime}
\end{figure}

\textbf{Efficiency.}
For tool installation, configuration, and learning time (Figure \ref{fig:ExpectScal+Eff+LearnTime}), practitioners are most satisfied when these activities take less than 10 minutes or between 10-60 minutes, with 26.8\% and 32.7\% favoring these ranges, respectively. Satisfaction decreases as time increases, with 22.3\% satisfied within 1-3 hours, 13.6\% for 3-24 hours, and the least satisfaction at 4.5\% for several days.
Regarding anomaly detection time, most practitioners (around 80\%) require detection to be within 1 minute. Specifically, 38.8\% are only satisfied with detection in less than 1 second, 51.3\% within 5 seconds, and 66\% within 20 seconds. Interestingly, 21.1\% are satisfied with detection taking longer (1-10 minutes). As one practitioner explained: ``\textit{Log anomalies do not necessarily need to be detected in real-time, and some anomalies may not cause program crashes immediately.}''.

\textbf{Scalability.}
Scalability explores the expectation of the maximum number of logs that can be processed and analyzed as the volume of logs grows while maintaining accurate and timely detection of anomalies. 
As shown in Figure~\ref{fig:ExpectScal+Eff+LearnTime}, 23.4\% of surveyed practitioners are satisfied with tools handling less than 10,000 logs. Satisfaction is 25.2\% for tools managing between 10,000-100,000 logs and 18.6\% for those handling 100,000-1,000,000 logs. The highest satisfaction, at 32.8\%, is for tools capable of handling over 1,000,000 logs.


\begin{center}
    \resizebox{\linewidth}{!}{
\begin{tabular}{l!{\vrule width 1pt}p{0.9\columnwidth}}
    \makecell{{\LARGE \faLightbulbO}}  &\textbf{Finding 8.} 
    More than half of the surveyed practitioners expect automated log anomaly detection tools to handle at least 100,000 logs, with installation, configuration, and learning time of less than 1 hour, and anomaly detection time to be under 5 seconds when an anomaly appears.
\end{tabular}}
\end{center}

\begin{figure}[h]
  \centering
  \includegraphics[width=\linewidth]{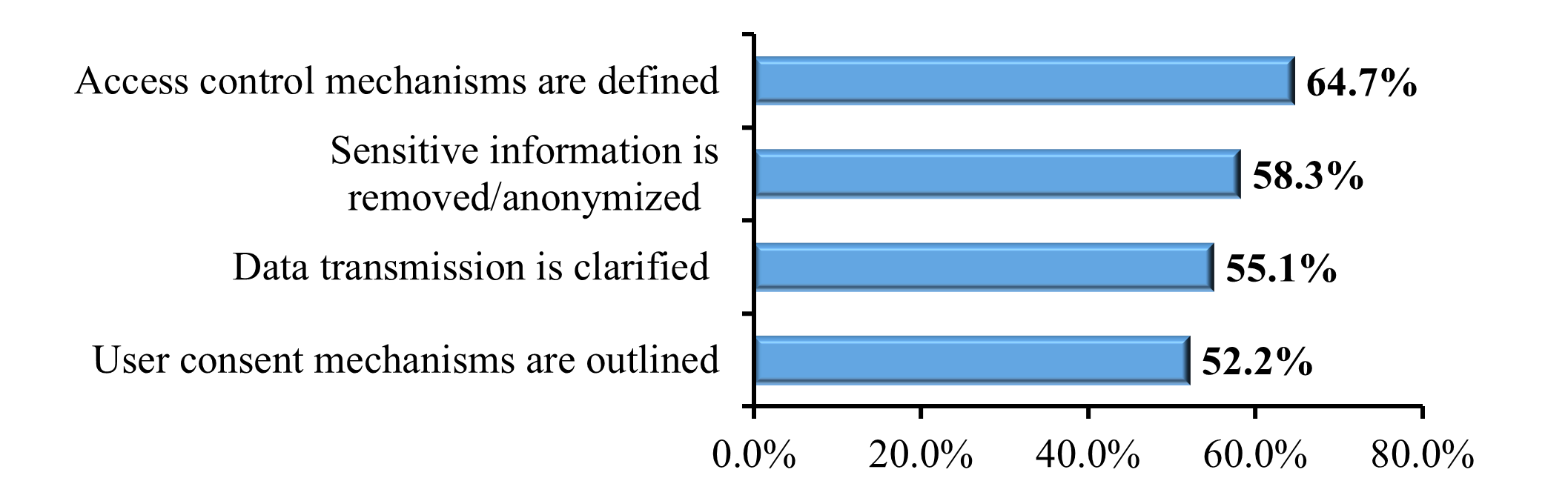}
  \captionsetup{skip=3pt}
  \caption{The surveyed practitioners' satisfaction in terms of privacy protection.}
  \label{fig:SecConcerns}
\end{figure}
\textbf{Privacy protection.}
We explore the expectations of the surveyed practitioners on privacy protection. As shown in Figure~\ref{fig:SecConcerns}, 64.7\% of practitioners are satisfied with automated log anomaly detection tools that clearly define access control mechanisms. Additionally, more than half of the practitioners accept these tools if privacy protection measures are explicitly stated, such as removing or anonymizing sensitive information (58.3\%), clarifying data transmission protocols like encryption (55.1\%), and transparently outlining user consent mechanisms for data collection and processing (52.2\%).

\subsection{RQ4: Gap Between the Current Research and Practitioners' Needs}\label{S:rq4}

\begin{table}[]
\scriptsize
\caption{The surveyed practitioners' capability expectations and the capabilities of current research. \textit{Sat. Rate} represents the satisfaction rate of surveyed practitioners choices. \textit{Interpret.}, \textit{General.}, \textit{Custom.}, and \textit{Priv.} represent interpretability, generalizability, customization, and privacy protection, respectively.}
\label{tab:lr}
\renewcommand\arraystretch{0.9}
\begin{tabular}{p{1.2cm}<{\centering}|p{1cm}<{\centering}|p{1cm}|p{3.3cm}<{\centering}}
\toprule
\textbf{Factor} & \multicolumn{2}{c|}{\textbf{Type}} & \textbf{Papers} \\\cline{1-4}
\multirow{5}{*}{\makecell{Data\\Resource}} & \multicolumn{2}{c|}{\makecell{Historical\\normal log}} & \cite{guo2024logformer}, \cite{hashemi2024onelog}, \cite{xie2024logsd}, \cite{zhang2024metalog}, \cite{zhang2023semi}, \cite{wang2023logonline}, \cite{huang2023twin}, \cite{lee2023heterogeneous}, \cite{xiao2023loader}, \cite{fu2023mlog}, \cite{huo2023autolog}, \cite{wang2023deepuserlog}, \cite{xie2023logrep}, \cite{hashemi2022sialog}, \cite{zhang2022deeptralog}, \cite{le2021log}, \cite{zhao2021empirical}, \cite{yang2021semi}, \cite{jia2021logflash}, \cite{chen2020logtransfer}, \cite{li2020swisslog}, \cite{zhang2019robust}, \cite{meng2019loganomaly}, \cite{du2017deeplog}, \cite{nedelkoski2020self}, \cite{wang2021multi}, \cite{zhang2022cat}, \cite{jia2017approach}, \cite{duan2023aclog}, \cite{yin2020improving}, \cite{wang2022maddc}, \cite{zhang2021log}\\\cline{2-4}
 & \multicolumn{2}{c|}{\makecell{Historical\\abnormal log}} & \cite{guo2024logformer}, \cite{hashemi2024onelog}, \cite{zhang2024metalog}, \cite{lee2023heterogeneous}, \cite{fu2023mlog}, \cite{huo2023autolog}, \cite{wang2023deepuserlog}, \cite{xie2023logrep}, \cite{hashemi2022sialog}, \cite{le2021log}, \cite{chen2020logtransfer}, \cite{li2020swisslog}, \cite{zhang2019robust}, \cite{duan2023aclog}, \cite{meng2021logclass} \\\cline{2-4}
 & \multicolumn{2}{c|}{\makecell{Historical\\unlabeled log}} & \cite{yang2023try}, \cite{farzad2020unsupervised}, \cite{kim2020automatic}, \cite{yang2021semi}, \cite{meng2021logclass}\\\cline{2-4}
 & \multicolumn{2}{c|}{Metric} & \cite{huang2023twin}, \cite{lee2023heterogeneous}, \cite{du2017deeplog}\\\cline{2-4}
 & \multicolumn{2}{c|}{Trace} & \cite{huang2023twin}, \cite{zhang2022deeptralog} \\\cline{1-4}
\multirow{4}{*}{Granularity} & \multicolumn{2}{c|}{Event} & \cite{hashemi2024onelog}, \cite{zhang2019robust}, \cite{nedelkoski2020self}, \cite{farzad2020unsupervised}, \cite{kim2020automatic}, \cite{meng2021logclass}\\\cline{2-4}
 & \multicolumn{2}{c|}{\makecell{Sequence\\(window size)}} & \cite{guo2024logformer}, \cite{hashemi2024onelog}, \cite{xie2024logsd}, \cite{yang2023try}, \cite{wang2023logonline}, \cite{huang2023twin}, \cite{fu2023mlog}, \cite{huo2023autolog}, \cite{wang2023deepuserlog}, \cite{xie2023logrep}, \cite{hashemi2022sialog}, \cite{le2021log}, \cite{zhao2021empirical}, \cite{yang2021semi}, \cite{chen2020logtransfer}, \cite{li2020swisslog}, \cite{meng2019loganomaly}, \cite{du2017deeplog}, \cite{farzad2020unsupervised}, \cite{wang2021multi}, \cite{duan2023aclog}, \cite{yin2020improving}, \cite{wang2022maddc}, \cite{zhang2021log}\\\cline{2-4}
 & \multicolumn{2}{c|}{\makecell{Sequence\\(time intervals)}} & \cite{yang2023try}, \cite{lee2023heterogeneous}, \cite{jia2021logflash}, \cite{zhang2022cat} \\\cline{2-4}
 & \multicolumn{2}{c|}{\makecell{Sequence\\(timestamp/session)}} & \cite{guo2024logformer}, \cite{hashemi2024onelog}, \cite{xie2024logsd}, \cite{zhang2024metalog}, \cite{yang2023try}, \cite{zhang2023semi}, \cite{wang2023logonline}, \cite{xiao2023loader}, \cite{fu2023mlog}, \cite{huo2023autolog}, \cite{wang2023deepuserlog}, \cite{xie2023logrep}, \cite{hashemi2022sialog}, \cite{zhang2022deeptralog}, \cite{le2021log}, \cite{yang2021semi}, \cite{li2020swisslog}, \cite{zhang2019robust}, \cite{meng2019loganomaly}, \cite{du2017deeplog}, \cite{wang2021multi}, \cite{zhang2022cat}, \cite{jia2017approach}, \cite{yin2020improving}, \cite{wang2022maddc}, \cite{zhang2021log}\\
 \toprule
\textbf{Factor} & \multicolumn{2}{c|}{\textbf{Sat. Rate}} & \textbf{Papers} \\\cline{1-4}
\multirow{8}{*}{Effectiveness} & \multirow{3}{*}{Recall} & 80-100\% & \cite{guo2024logformer}, \cite{hashemi2024onelog}, \cite{xie2024logsd}, \cite{zhang2024metalog}, \cite{yang2023try}, \cite{zhang2023semi}, \cite{wang2023logonline}, \cite{huang2023twin}, \cite{lee2023heterogeneous}, \cite{xiao2023loader}, \cite{fu2023mlog}, \cite{huo2023autolog}, \cite{wang2023deepuserlog}, \cite{xie2023logrep}, \cite{hashemi2022sialog}, \cite{zhang2022deeptralog}, \cite{le2021log}, \cite{yang2021semi}, \cite{jia2021logflash}, \cite{chen2020logtransfer}, \cite{li2020swisslog}, \cite{zhang2019robust}, \cite{meng2019loganomaly}, \cite{du2017deeplog}, \cite{farzad2020unsupervised}, \cite{zhang2022cat}, \cite{duan2023aclog}, \cite{yin2020improving}, \cite{wang2022maddc}, \cite{zhang2021log}\\\cline{3-4}
 &  & 60-80\% & \cite{wang2021multi}, \cite{jia2017approach}\\\cline{3-4}
 &  & 40-60\% & \cite{nedelkoski2020self} \\\cline{3-4}
 &  & <40\% & - \\\cline{3-4}
 &  & ? & \cite{zhao2021empirical}\footnotemark[1]{}, \cite{kim2020automatic}\footnotemark[2]{}, , \cite{meng2021logclass}\footnotemark[1]{}\\\cline{2-4}
 & \multirow{5}{*}{Precision} & 80-100\% & \cite{guo2024logformer}, \cite{hashemi2024onelog}, \cite{xie2024logsd}, \cite{yang2023try}, \cite{zhang2023semi}, \cite{huang2023twin}, \cite{lee2023heterogeneous}, \cite{xiao2023loader}, \cite{fu2023mlog}, \cite{huo2023autolog}, \cite{wang2023deepuserlog}, \cite{hashemi2022sialog}, \cite{zhang2022deeptralog}, \cite{le2021log}, \cite{yang2021semi}, \cite{jia2021logflash}, \cite{chen2020logtransfer}, \cite{li2020swisslog}, \cite{meng2019loganomaly}, \cite{du2017deeplog}, \cite{farzad2020unsupervised}, \cite{zhang2022cat}, \cite{jia2017approach}, \cite{yin2020improving}, \cite{wang2022maddc}, \cite{zhang2021log}\\\cline{3-4}
 &  & 60-80\% & \cite{wang2023logonline}, \cite{xie2023logrep}, \cite{zhang2019robust}, \cite{nedelkoski2020self}, \cite{wang2021multi}, \cite{duan2023aclog}\\\cline{3-4}
 &  & 40-60\% & - \\\cline{3-4}
 &  & <40\% & \cite{zhang2024metalog} \\\cline{3-4}
 &  & ? & \cite{zhao2021empirical}\footnotemark[1]{}, \cite{kim2020automatic}\footnotemark[2]{}, \cite{meng2021logclass}\footnotemark[1]{}\\\cline{1-4}
\multirow{5}{*}{Efficiency} & \multicolumn{2}{c|}{<1 millisecond} & \cite{guo2024logformer}, \cite{zhang2024metalog}, \cite{zhang2022deeptralog}, \cite{yang2021semi}, \cite{jia2021logflash}, \cite{du2017deeplog}, \cite{meng2021logclass}\\\cline{2-4}
& \multicolumn{2}{c|}{1 millisecond-1 second} & \cite{xie2024logsd}, \cite{yang2023try}, \cite{huo2023autolog}, \cite{farzad2020unsupervised}, \cite{kim2020automatic} \\\cline{2-4}
& \multicolumn{2}{c|}{1-5 second} & \cite{huang2023twin} \\\cline{2-4}
& \multicolumn{2}{c|}{>5 seconds} & - \\\cline{2-4}
& \multicolumn{2}{c|}{?} & \cite{hashemi2024onelog}, \cite{zhang2023semi}, \cite{wang2023logonline}, \cite{lee2023heterogeneous}, \cite{xiao2023loader}, \cite{fu2023mlog}, \cite{wang2023deepuserlog}, \cite{xie2023logrep}, \cite{hashemi2022sialog}, \cite{le2021log}, \cite{zhao2021empirical}, \cite{chen2020logtransfer}, \cite{li2020swisslog}, \cite{zhang2019robust}, \cite{meng2019loganomaly}, \cite{nedelkoski2020self}, \cite{wang2021multi}, \cite{zhang2022cat}, \cite{jia2017approach}, \cite{duan2023aclog}, \cite{yin2020improving}, \cite{wang2022maddc}, \cite{zhang2021log}\\\cline{1-4}
\multirow{2}{*}{Scalability} & \multicolumn{2}{c|}{$\leq$1,000,000} & \cite{kim2020automatic} \\\cline{2-4}
& \multicolumn{2}{c|}{>1,000,000} & \cite{guo2024logformer}, \cite{hashemi2024onelog}, \cite{xie2024logsd}, \cite{zhang2024metalog}, \cite{yang2023try}, \cite{zhang2023semi}, \cite{wang2023logonline}, \cite{huang2023twin}, \cite{lee2023heterogeneous}, \cite{xiao2023loader}, \cite{fu2023mlog}, \cite{huo2023autolog}, \cite{wang2023deepuserlog}, \cite{xie2023logrep}, \cite{hashemi2022sialog}, \cite{zhang2022deeptralog}, \cite{le2021log}, \cite{zhao2021empirical}, \cite{yang2021semi}, \cite{jia2021logflash}, \cite{chen2020logtransfer}, \cite{li2020swisslog}, \cite{zhang2019robust}, \cite{meng2019loganomaly}, \cite{du2017deeplog}, \cite{nedelkoski2020self}, \cite{farzad2020unsupervised}, \cite{wang2021multi}, \cite{zhang2022cat}, \cite{jia2017approach}, \cite{duan2023aclog}, \cite{yin2020improving}, \cite{wang2022maddc}, \cite{meng2021logclass}, \cite{zhang2021log}\\
\toprule
\textbf{Factor} & \multicolumn{2}{c|}{\textbf{Support}} & \textbf{Papers} \\\cline{1-4}
Interpret. & \multicolumn{2}{c|}{Yes} & \cite{zhao2021empirical} \\\cline{1-4}
General. & \multicolumn{2}{c|}{\makecell{Cross-project}} & \cite{guo2024logformer}, \cite{hashemi2024onelog}, \cite{zhang2024metalog}, \cite{chen2020logtransfer} \\\cline{1-4}
Custom. & \multicolumn{2}{c|}{Yes} & - \\\cline{1-4}
Priv. & \multicolumn{2}{c|}{Yes} & - \\\bottomrule
\end{tabular}
\end{table}
\footnotetext[1]{The technique proposed in the paper uses F1 for evaluation, and its recall and precision are likely to be in the 80-100\% range.}
\footnotetext[2]{The technique proposed in the paper uses the Debugging Effectiveness score for evaluation, and its recall and precision cannot be estimated.}

After our literature review process, we identify a total of 36 papers. Table~\ref{tab:lr} shows the capabilities of state-of-the-art log anomaly detection techniques in terms of nine factors.

\textbf{Data Resource.} 
From Table~\ref{tab:lr}, most of these papers focus solely on log data for detecting anomalies. Specifically, 32 papers (88.9\%) use historical labeled normal data, 15 papers (41.7\%) use historical labeled abnormal data, and 3 papers \cite{yang2023try, farzad2020unsupervised, kim2020automatic} (8.3\%) only rely on historical unlabeled data for training. 
However, our survey results reveal that only 39.3\% and 43.5\% of practitioners always have access to historical labeled normal and abnormal data, respectively. Thus, there is a noticeable gap between the availability of historical labeled data and its actual usage in real-world scenarios. 
Furthermore, only 4 papers (11.1\%) integrate other types of data, such as metrics and traces, to aid in anomaly detection. Surprisingly, our survey shows that 83.7\% and 74.9\% of practitioners find metrics and traces always or sometimes available, respectively. Despite this availability, these data types are not utilized in the majority of studies. As one surveyed practitioner aptly stated: ``\textit{Different types of data are essential to train an effective anomaly log detection tool.}''. This underscores the importance of considering these additional data types for anomaly detection.

\textbf{Granularity level.}
Most of the papers evaluate their techniques on two public datasets, Hadoop Distributed File System (HDFS) \cite{xu2009detecting} and Blue Gene/L supercomputer (BGL) \cite{oliner2007supercomputers}, typically grouping logs into sequences using sessions and window sizes. Consequently, 26 (72.2\%) and 24 (66.7\%) studies operate at the log sequence level based on these grouping strategies, respectively. Additionally, 4 studies (11.1\%) group logs according to time intervals. While 6 studies (16.7\%) perform log event level anomaly detection, 3 of them \cite{nedelkoski2020self, kim2020automatic, meng2021logclass} only work at the event level. 
Our survey results show that the top preference of practitioners surveyed (30.1\%) is to work at the log sequence level based on window size, while 29.5\% prefer to work at the log event level, which is the second preference.

\begin{center}
    \resizebox{\linewidth}{!}{
\begin{tabular}{l!{\vrule width 1pt}p{0.9\columnwidth}}
    \makecell{{\LARGE \faLightbulbO}}  &\textbf{Finding 9.} 
    The majority (89\%) of the studies primarily rely on historical labeled normal log data for anomaly detection, which contrasts sharply with the surveyed data availability (39.3\%). Few studies have incorporated other types of data for anomaly detection, which practitioners deem important. Additionally, a few studies work at the log event level, while 29.5\% of surveyed practitioners prefer this granularity.
\end{tabular}}
\end{center} 

\textbf{Effectiveness.} 
The most important factor identified in the survey is the correct detection rates, measured by recall and precision. 
As demonstrated in Table~\ref{tab:lr}, 30 papers (83.3\%) and 26 (72.2\%) achieve recall and precision in the range of 80\%-100\%, respectively, meeting the requirements of all surveyed practitioners. 32 papers (88.9\%) achieve recall and precision within 60\%-80\%, which can satisfy at least 60.9\% and 57.7\% of the surveyed practitioners.
We categorize some papers as ``?'' since we cannot ascertain the detection rates of the log anomaly detection techniques they presented.

\textbf{Efficiency.} 
Regarding the installation, configuration, and learning
time costs, our survey results reveal that an automated log anomaly detection technique with capabilities in the ranges of <10 minutes, 10-60 minutes, 1-3 hours, and 3-24 hours satisfies 100\%, at least 73.1\%, 40.4\%, and 18.1\% of the surveyed practitioners, respectively. As one surveyed practitioner mentioned: ``\textit{If I find these tools and friendly install these tools, I will use these tools.}''. However, none of the reviewed papers utilize this metric to evaluate tool efficiency. Regarding the time taken to detect anomalies, our survey results indicate that an automated log anomaly detection technique capable of detecting anomalies within 1 second can satisfy 100\% of the surveyed practitioners, while a detection time within 5 seconds can satisfy at least 61.2\% of them. As shown in Table~\ref{tab:lr}, 12 papers (33.3\%) meet the 1-second requirement, and 13 papers (36.1\%) meet the 5-second requirement. Notably, 7 papers demonstrate detection times within 1 millisecond. However, over half of the papers (63.9\%) do not describe the testing time of their proposed techniques and thus are categorized as ``?''.

\textbf{Scalability.} 
Our survey results point out that an automated log anomaly detection technique handling at least 1,000,000 logs satisfies 100\% of the surveyed practitioners. Table~\ref{tab:lr} shows that 35 papers (i.e., excluding \cite{kim2020automatic}) satisfy all surveyed practitioners, as they have evaluated their proposed techniques on at least one public dataset containing more than 1,000,000 logs.

\textbf{Interpretability.}
Most log anomaly detection techniques solely determine whether a testing log is anomalous or not. However, 84.5\% surveyed practitioners desire more than just a binary classification; they seek a rationale for detected anomalies to aid in system maintenance, allowing for the adoption of appropriate actions.
As a practitioner emphasized: ``\textit{The tool's ability to provide the rationale for a log being flagged as an anomaly is critical to understanding the underlying issue and taking appropriate action. Without this explanation, we have no way of knowing what happened, why, and how to fix it.}''.
Among the surveyed papers, only Zhao et al. \cite{zhao2021empirical} provide an interpretable report for anomalies. 
This report comprises detailed time series information and a workflow of task execution. The authors aim to leverage domain knowledge and human-computer interactive log analysis to facilitate intuitive anomaly detection. However, as there are no results reporting human evaluation, it remains unclear whether the interpretability of the report is indeed helpful.

\textbf{Generalizability.}
As training on a single dataset and testing on the same dataset typically result in a model learning a simplistic log structure, we explore studies to determine if they have conducted cross-project evaluations to demonstrate a degree of generalizability. According to our survey results, 83.8\% of practitioners expect log anomaly detection techniques to possess generalizability, enabling them to effectively analyze logs with diverse structures or characteristics. However, as indicated in Table~\ref{tab:lr}, only 4 papers \cite{guo2024logformer, hashemi2024onelog, zhang2024metalog, chen2020logtransfer} evaluate their techniques across different domains, training on a source domain and testing on a different target domain. While 2 papers \cite{guo2024logformer,chen2020logtransfer} achieve satisfactory results on public datasets like HDFS, these datasets have limited log information, addressing only one or two types of abnormal patterns. For instance, BGL logs primarily contain information related to the RAS kernel, while HDFS logs describe operations on the storage block pool \cite{zhang2024metalog}. Moreover, publicly available data is typically single-sourced, and industrial datasets are often access-restricted due to security and privacy concerns, as highlighted by Lee et al. \cite{lee2023heterogeneous}. Mäntylä \cite{hashemi2024onelog} also claim that achieving high accuracy is possible only when the target domain's data is sufficiently similar to the source domain's. Consequently, the lack of diverse log data and limited cross-project evaluation present substantial challenges for achieving generalizability.

\textbf{Customization and privacy protection.}
None of the log anomaly detection techniques proposed in the 36 papers we reviewed offer customization or privacy protection. These factors, which are crucial to most of the surveyed practitioners, have been overlooked in these studies.
As some practitioners noted: ``\textit{Some exceptions are actively thrown out by the program, and some exceptions are real anomalies. How to determine whether they are actively thrown out for debug analysis is very important. That is, the tool should allow users to have different customization requirements.}'',
``\textit{Log data often includes user identifiers and IP addresses. Without proper privacy safeguards, our information could be exposed, so I would be concerned about this tool invading my privacy.}''.

\begin{center}
    \resizebox{\linewidth}{!}{
\begin{tabular}{l!{\vrule width 1pt}p{0.9\columnwidth}}
    \makecell{{\LARGE \faLightbulbO}}  &\textbf{Finding 10.} 
    Most (72\%) of the automated log anomaly detection techniques proposed in the surveyed papers achieve recall and precision within 80\%-100\%, meeting the expectations of all practitioners. One-third of the techniques can detect anomalies within one second, satisfying all practitioners' expectations for anomaly detection time. However, few or no papers address practitioners' needs for interpretability, generalizability, customization, and privacy protection.
\end{tabular}}
\end{center} 

\textbf{Challenges of adopting existing log anomaly detection techniques.}
Although most of the state-of-the-art log anomaly detection techniques reviewed in our study achieve high accurate detection rates (i.e., recall and precision within 80\%-100\%), around half of the surveyed practitioners (51\%) do not prioritize these techniques due to various concerns expressed at the end of the survey. We categorize three main types of concerns and
present their frequencies using the multiplication symbol (\textit{X}).
\textbf{\textit{(a) Lack of interpretability (26X)}:} Many practitioners note that existing techniques often provide only binary results without explanations, which they find insufficient. ``\textit{I will consider using these models if they provide hints for me to debug my system.}'' and ``\textit{The models in research are not a priority because the unexplainability introduced by deep learning algorithms cannot explain its predictions.}''.
\textbf{\textit{(b) Unsure capability of handling various log data (19X)}:} Some practitioners point out that existing techniques are trained on limited datasets that may not represent industrial data adequately. ``\textit{I will not use these models. The currently used datasets are still relatively one-sided and cannot widely represent the distribution of data sets in the industry.}''.
\textbf{\textit{(c) Lack of user-friendly usage (12X)}:} Several surveyed practitioners express concerns about the complexity of techniques and whether they can be used easily. ``\textit{The existing log anomaly detection techniques often require expertise in model tuning, making them too challenging for me to try.}''.


%% file: Discussion.tex
\subsection{Implications}
Our results highlight key implications for research and industrial communities:

\textbf{(1) Large improvement in interpretability and generalizability of existing techniques is needed.}
To achieve at least an 80\% satisfaction rate, these techniques for automated log anomaly detection need to provide a rationale for their detection and efficiently detect various log anomalies. 
However, our literature review reveals that the majority of the surveyed techniques overlook these factors. Notably, only one paper offers an interpretable report for detected anomalies.
While some practitioners emphasize the importance of generalizability for adopting these techniques, only four papers evaluate the generalizability of their techniques across different projects. However, the effectiveness of these techniques remains unclear or unsatisfactory due to limited datasets and single log structure. Thus, we encourage researchers to consider providing interpretable results for detected anomalies and developing techniques capable of handling diverse log data.

\textbf{(2) Community-wide effort to integrate state-of-the-art techniques into automated log anomaly detection tools with customization is needed.} Through our survey, the majority of practitioners (78.2\%) prioritize customization in automated log anomaly detection tools, such as adapting anomaly alert thresholds, as detailed in Section~\ref{S:rq3}. Nevertheless, none of the techniques proposed in the surveyed papers discuss this issue. To bridge this gap, a collective effort from the community is essential to develop automated tools that offer robust customization options that better meet industry needs. 
``\textit{We can consider using the automated tools. However, given the complexity of our business, we are uncertain about their ability to adapt to our specific task of identifying real anomalies, such as an alert threshold.}''.

\textbf{(3) Large demand for user-friendly automated tools is needed.} Ease of use is a major concern among surveyed practitioners, with at least 76.6\% considering it important (Section~\ref{S:rq3}). This includes aspects such as the tool's installation and deployment, with only 18\% willing to invest more than three hours in these tasks. Therefore, there is a clear need for the development of easy-to-use automated tools.
``\textit{If the automated log anomaly detection tool lacks a user interface, I may not consider it, as convenience is key for me. I prefer tools that are readily available and easy to use.}''.
Moreover, over half of the surveyed practitioners stress compatibility issues when using log monitoring tools for log anomaly detection. 
``\textit{The automated tool should feature a modular architecture to ensure easy integration into our existing systems. Alternatively, providing APIs to facilitate integration with diverse legacy systems would greatly benefit us, enabling smooth data exchange and enhancing overall interoperability.}''. Consequently, prioritizing the development of user-friendly automated log anomaly detection tools that are easy to use and compatible is essential.

\textbf{(4) Suggestions for LLM-based automated log anomaly detection.} 
We have encouraged practitioners to share their insights on the potential of LLMs to aid in log anomaly detection, given their increasing popularity for software engineering tasks \cite{wang2024software,du2024evaluating,fan2023large}. Surveyed practitioners express varied opinions, which we summarize for further exploration:
(a) Integrating LLMs as a plug-in function to improve interpretability: ``\textit{It should plug in with my existing tool instead of being an add-on tool to provide an explanation.}''
(b) Utilizing LLMs' prompt learning with high-quality data: ``\textit{The accuracy of your tool to detect anomalies in the industry is very, very important. Even one false anomaly detection may have serious consequences. Additionally, I'm worried about whether the training time costs will end up being passed on to us users. While prompt learning could avoid the need for fine-tuning, the data used for prompting needs to be carefully chosen because it has a significant impact on LLM performance.}''
(c) Addressing data security challenges for in-context learning: ``\textit{My main concern about using LLMs is data security. If I use in-context learning to prompt LLMs, will the inference process of LLMs expose my data? Are there sufficient safeguards in place to protect my data?}''.

\subsection{Threats to Validity}
One of the potential threats to the validity of our survey is the possibility that some practitioners may not fully comprehend the questions. For example, some practitioners may not understand log anomaly detection and instead rely solely on automated tools to manage logs. Therefore, they may be unfamiliar with the questions about log anomaly detection. To mitigate this threat, we provided high-level instructions accompanied by a flow chart to clarify the questions. Furthermore, responses indicating ``I don't know'' will be excluded from the analysis. This threat to validity is common and considered tolerable in previous studies \cite{wang2023practitioners, yang2017language}.
Another potential threat to the validity of our study is that our sample of practitioners does not encompass the entire population of software engineers. Specifically, our practitioners are limited to those working for various companies and contributors to open-source projects hosted on GitHub in diverse roles. Consequently, our findings may not fully represent the expectations of all software engineers. For instance, our survey excludes practitioners who are not proficient in either English or Chinese. While we focus on several factors that may influence the adoption of automated log anomaly detection tools, there may be additional factors that our study has not addressed. We plan to investigate these factors in future research.

%% file: Related_Work.tex
\subsection{Automated Log Anomaly Detection}
Numerous machine learning and deep learning approaches have been proposed for automated log anomaly detection, including supervised, semi-supervised, and unsupervised approaches. 
Supervised approaches, leveraging abundant labeled data, typically outperform semi-supervised and unsupervised methods. Liang et al. \cite{liang2007failure} introduced four classifiers for predicting failure log events.  
Zhang et al. \cite{zhang2016automated} represented log templates as vectors combined with log template extraction with $tf-idf$, employing an LSTM model for anomaly prediction. Similarly, Vinayakumar et al. \cite{vinayakumar2017long} utilized a stacked-LSTM model to learn temporal patterns with sparse representations.  Lu et al. \cite{lu2018detecting} learned local semantic information from log data utilizing CNN with three filters. 
Zhang et al. \cite{zhang2019robust} integrated attention mechanisms with a Bi-LSTM model to capture comprehensive log sequences bidirectionally. Le et al. \cite{le2021log} leveraged BERT for log representation and a Transformer encoder for log anomaly detection.
Semi-supervised and unsupervised approaches, requiring less labeled data, are considered more practical. 
For instance, Du et al. \cite{du2017deeplog} autonomously learned log patterns with an LSTM model and detected anomalies based on deviations from normal execution. Meng et al. \cite{meng2019loganomaly} incorporated semantic information by matching log sequences against generated templates to learn normal patterns. Yang et al. \cite{yang2021semi} combined HDBSCAN clustering for probabilistic label estimation with an attention-based GRU model. Huo et al. \cite{huo2023autolog} examined logging statements and constructed execution graphs to identify log-related paths. Anomaly labels are then propagated to each execution path based on the analysis of labeled sequence anomalies.
Lin et al. \cite{lin2016log} proposed an unsupervised approach that clusters log sequences hierarchically, taking into account the weights of log events. 
Lee et al. \cite{lee2023lanobert} learned log information autonomously without parsing based on BERT. 
Yang et al. \cite{yang2023try} incorporated a lightweight semantic-based log representation in traditional unsupervised principal component analysis for log anomaly detection. 


\subsection{Studies on Log Analysis}
Several recent studies have investigated log analysis through empirical methods. For instance, Li et al. \cite{li2023they} employed semi-structured interviews and surveys to examine practitioners' expectations regarding the readability of log messages and to explore the potential for automatically classifying the readability of these messages. Yang et al.  \cite{yang2023interview} conducted interviews to understand how developers utilize logs within an embedded software engineering context. Rong et al. \cite{rong2023developers} carried out an empirical study using mixed methods, including questionnaire surveys, semi-structured interviews, and code analyses, to explore the relationships between developers’ profiles, experiences, and their logging practices. He et al. \cite{he2021survey} examined the four main steps in the automated log analysis framework: logging, log compression, log parsing, and log mining. Fu et al. \cite{fu2014developers} surveyed 54 experienced developers at Microsoft to investigate logging practices, particularly where developers choose to log. However, there is a lack of research focused on the practices, issues, and expectations of practitioners regarding automated log anomaly detection tools.

%% file: Conclusion.tex
In this paper, we interview 15 professionals and survey 312 practitioners about their log anomaly detection practices, the issues they face, and their expectations for automated log anomaly detection tools. Practitioners express dissatisfaction with existing log monitoring tools for log anomaly detection and indicate a willingness to adopt automated log anomaly detection tools if certain aspects are satisfied, including granularity level, evaluation metrics, effectiveness, efficiency, scalability, and privacy protection. We also compare the capabilities of current research with practitioners' expectations through a literature review, offering insights for further improvements to more effectively meet practitioners' needs.